\documentclass[preprint,pra,superscriptaddress,amsmath,amssymb]{revtex4-1}
\usepackage{amsmath}
\usepackage{amsfonts}
\usepackage{amssymb}
\usepackage{graphicx}
\usepackage{color}
\usepackage{txfonts}
\usepackage[titletoc]{appendix}
\usepackage[colorlinks={true}]{hyperref}
\hypersetup{citecolor={blue}, filecolor={blue}, linkcolor={blue}, urlcolor={blue}}

\begin{document}
\preprint{}
\title{Cyclic three-level-pulse-area theorem for enantioselective  state transfer of chiral
molecules}
\author{Yu Guo}
%\email{guoyu@csust.edu.cn}
\affiliation{Hunan Key Laboratory of Nanophotonics and Devices, School of Physics and Electronics, Central South University,
Changsha 410083, China}
\affiliation{Hunan Provincial Key Laboratory of Flexible Electronic Materials Genome Engineering, School
of Physics and Electronic Science, Changsha University of Science
and Technology, Changsha 410114, China}
\affiliation{The Key Laboratory of Low-Dimensional Quantum Structures and Quantum Control of Ministry of Education, Hunan Normal University,
Changsha 410081, China}
\author{Xun Gong}
%dgisme@foxmail.com
\affiliation{Hunan Provincial Key Laboratory of Flexible Electronic Materials Genome Engineering, School
of Physics and Electronic Science, Changsha University of Science
and Technology, Changsha 410114, China}
\author{Songshan Ma}
%songshan@csu.edu.cn
\affiliation{Hunan Key Laboratory of Nanophotonics and Devices, School of Physics and Electronics, Central South University,
Changsha 410083, China}
\author{Chuan-Cun Shu}
\email{cc.shu@csu.edu.cn he/him/his}
\affiliation{Hunan Key Laboratory of Nanophotonics and Devices, School of Physics and Electronics, Central South University,
Changsha 410083, China}
\begin{abstract}
We derive a  pulse-area theorem for a cyclic three-level system, an archetypal model for exploring enantioselective state transfer (ESST) in chiral molecules driven by three linearly polarized microwave pulses. By dividing the closed-loop excitation into two separate stages, we obtain both amplitude and phase conditions of three control fields to generate high fidelity of  ESST. As a proof of principle, we apply this pulse-area theorem to the cyclohexylmethanol molecules ($\text{C}_{7}\text{H}_{14}\text{O}$), for which three rotational states are connected by the $a$-type, $b$-type, and $c$-type components of the transition dipole moments in both center-frequency resonant and detuned conditions. Our results show that two enantiomers with opposite handedness can be transferred to different target states by designing three microwave pulses that satisfy the amplitude and phase conditions at the transition frequencies. The corresponding control schemes are robust against the time delays between the two stages. We suggest that the two control fields used in the second stage should be applied simultaneously for practical applications. This work contributes an alternative pulse-area theorem to the field of quantum control, which has the potential to determine the chirality of enantiomers in a mixture.
\end{abstract}
\maketitle
\section{Introduction}
Since Louis Pasteur first reported molecular chirality in 1848 \cite{Pasteur1848}, the theoretical and experimental study of chiral molecules has drawn increasing interest because of its fundamental importance in modern chemical and biochemical industries as well as quantum science \cite{Quack2008,review1,review2,review3}. Two enantiomers of chiral molecules with opposite handedness have the same components and configuration for spatial reflection. It implies that distinguishing enantiomers from each other, highly related to the discrimination, separation, and purification of chiral molecules, remains a formidable task by comparing general physical and chemical properties, such as boiling points, melting points, and densities. Based on chemical mechanisms and enantiomer-specific interactions with auxiliary substances, many spectroscopic techniques were established to detect enantiomers of chiral molecules with different handedness \cite{Vogt,Hazen2003,Patterson2014}. The traditional techniques, such as crystallization and chiral chromatography, are complicated and expensive and require significantly more protracted than seconds, leading to discrimination of chiral molecules out of reach.\\ \indent
By taking advantage of the sign difference property of optical rotations, it has become promising to select enantiomers by designing coherent quantum optical schemes \cite{Heppke1999,Brumer2001,Thanopulos2003,Frishman2004,Li2007,Li2008,Ye2018,Ye2019,Yachmenev2019,Wu2020,Torosov2020,Torosov2020-2,Ye2020,Wu2020-OE,Chen2020,Xu2020,liyong2021,Tutunnikov2021}. The concept of the adiabatic passage techniques, such as stimulated Raman adiabatic passages \cite{stirap1,stirap2} and shortcuts to adiabaticity \cite{sta1,sta2}, was proposed to generate efficient and robust detection and separation of chiral molecules \cite{Shapiro2000,Kral2001,Kral2003,Gerbasi2001,Vitanov2019,Wu2019}.  In order to meet adiabatic criteria,  the adiabatic passage techniques involve strict limitations on the control fields, and therefore, the corresponding control processes are usually slow and complicated. To that end, nonadiabatic schemes using much shorter durations of control pulses than the adiabatic approaches have been proposed to reach fast enantioselective excitation of chiral molecules \cite{Frishman2003,Hamedi2019}.  Experimentally, it has been demonstrated by using resonant microwave three-wave mixing (M3WM) techniques \cite{M3WM0,M3WM1,M3WM2,M3WM3,M3WM4,M3WM5,M3WM6,M3WM7}.  A common feature of both adiabatic and nonadiabatic control schemes  usually involves a closed-loop quantum system by cyclic coupling of three molecular (i.e., rotational or rovibrational) states (as shown in Fig. \ref{fig1}), which are resonantly driven through the $a$-type, $b$-type, and $c$-type components of the transition dipole moments by a  combination of three orthogonally polarized and phase-controlled microwave fields \cite{Leibscher2019}.  Since one of three cyclic couplings differs in the sign of the transition dipole moment in two opposite enantiomers, direct one-photon transition path from the ground state to a given target state constructively or destructively interferes with indirect two-photon transition path through an intermediate state, leading to enantioselective state transfer (ESST). Although the pulse areas of the three control fields that can generate enantioselective excitation have been experimentally examined in M3WM experiments, there is still lacking
a general pulse-areas theorem that can be used to directly calculate the amplitudes and phases of three time-dependent control pulses so as  to gain insights into the underlying coherent quantum control mechanism.\\ \indent
In this paper, we focus on ESST and present a three-level pulse-areas theorem analysis. Previous works had derived the pulse area theorems for the three-level quantum systems with the ladder-, $\Lambda$- and $V$-type configurations \cite{Sugny1,Guo2008,Shchedrin2015,shupra2019}, leading to many successful applications in coherent quantum control simulations and experiments \cite{Sola2018,jpca2019,shuprl2019,shupra2020,shuol2020,shupra2021}. Here, we take a strategy by dividing  the closed-loop three-level excitation into two separated stages, i.e.,  combining a two-level excitation and a time-delayed-open-loop three-level transition. We obtain a pulse-areas theorem of the three-level system with a $\Delta$-type configuration without applying the rotating-wave approximation. The derived pulse-areas theorem can calculate the exact amplitude and phase conditions for generating efficient ESST to the desired quantum state, which is examined in cyclohexylmethanol molecules ($\text{C}_{7}\text{H}_{14}\text{O}$) with different pulse sequences.   This work provides an essential reference for coherent control of ESST using closed-loop three-level interaction schemes. \\ \indent
The remainder of this paper is organized as follows.  In Sec. \ref{methods}, we describe the theoretical methods for obtaining a three-level pulse area theorem with cyclic coupling. We perform the numerical simulations to examine the derived pulse-areas theorem in Sec. \ref{simulation}. Finally, we conclude with a summary in Sec. \ref{con}.
\section{Theoretical method and numerical model}\label{methods}
\subsection{Closed-loop three-level model}
\begin{figure}
\centering \includegraphics[scale=0.4]{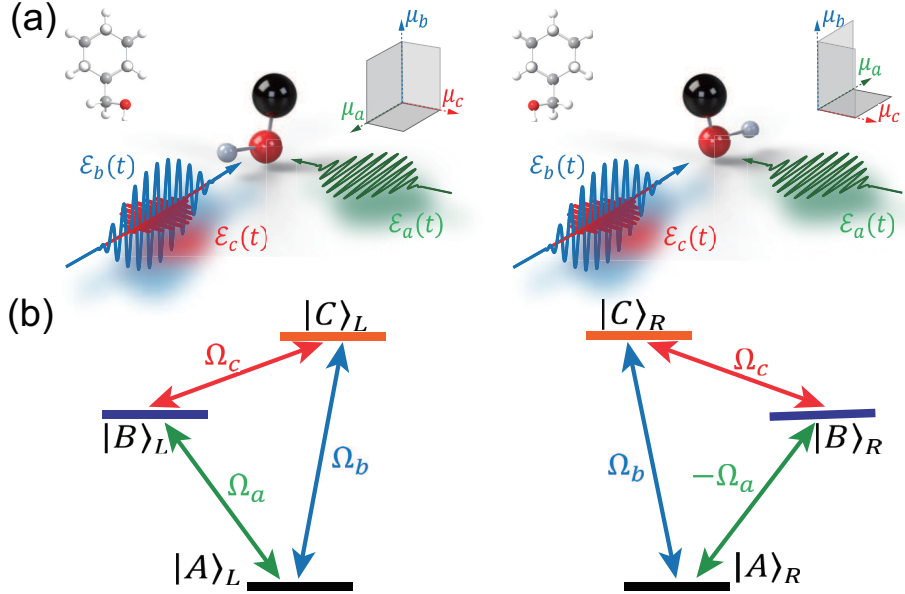} \caption{Schematic illustration of enantioselective  state transfer (ESST). (a) The space orientations of the transition dipole moments $\mu_{a/b/c}$ with respect to the polarization directions of three linearly polarized microwave pulses $\mathcal{E}_{a/b/c}(t)$ . (b) The corresponding closed-loop transitions within three rotational states $|A\rangle$, $|B\rangle$ and $|C\rangle$. The transition frequencies and couplings between states are identical, except for a difference due to the sign difference in $\mu_a$ The parameters of the cyclohexylmethanol molecules ($\text{C}_{7}\text{H}_{14}\text{O}$) are used for numerical simulations. }
\label{fig1}
\end{figure}
To describe our model,  three rotational states of asymmetric top  molecules, as shown in Fig. \ref{fig1},  are labeled as $|A\rangle$, $|B\rangle$ and $|C\rangle$ with a subscript $L/R$ for the left/right-handed enantiomer. The energies $E_A$, $E_B$ and $E_C$ of three rotational states  are identical for enantiomers. We mark three microwave control fields as  $\overrightarrow{\mathcal{E}}_{a/b/c}\left(t\right)$, which drive the three states with the transition dipole moments $\overrightarrow{\mu}_a$,  $\overrightarrow{\mu}_b$ and $\overrightarrow{\mu}_c$. As demonstrated in Refs. \cite{Leibscher2019,kevin2018} , the ESST scheme with the use of linearly polarized control fields requires that a combination of control fields $\overrightarrow{\mathcal{E}}_a$, $\overrightarrow{\mathcal{E}}_b$ and $\overrightarrow{\mathcal{E}}_c$ with three orthogonal polarization directions along the directions of  three dipole moment components $\overrightarrow{\mu}_a$,  $\overrightarrow{\mu}_b$ and $\overrightarrow{\mu}_c$.  To that end, we describe the time- and polarization-dependent electric fields  of the three control fields by
\begin{equation}
\overrightarrow{\mathcal{E}}_{a/b/c}\left(t\right)=\overrightarrow{e}_{a/b/c}\mathcal{E}_{a/b/c}f_{a/b/c}\left(t\right)\cos\left[\omega_{a/b/c}(t-t_{a/b/c})+\phi_{a/b/c}\right]
\end{equation}
where $\overrightarrow{e}_{a/b/c}$, $\mathcal{E}_{a/b/c}$, $f_{a/b/c}\left(t\right)$, $\omega_{a/b/c}$, $t_{a/b/c}$, and $\phi_{a/b/c}$ denote the polarization direction, the strength, envelope function, center frequency, center time,  and phase of $\overrightarrow{\mathcal{E}}_{a/b/c}\left(t\right)$, respectively. The triple product $\overrightarrow{\mu}_a\cdot(\overrightarrow{\mu}_b\times\overrightarrow{\mu}_c)$ is independent on the choice of the inertia principle axes $a$, $b$, and $c$,  but is of opposite sign for the left- and right-handed enantiomers.  For convenience, we specify that $\overrightarrow{\mu}_b$ and $\overrightarrow{\mu}_c$ are identical for two enantiomers, whereas $\overrightarrow{\mu}_a$  changes sign with the handedness. Thus, the Hamiltonian of two different handed-enantiomers in the presence of the control fields can be written as \begin{eqnarray}
H^{L,R}\left(t\right)&=&\left(\begin{array}{ccc}
E_A & \pm \Omega_a\left(t\right) & \Omega_b\left(t\right)\\
\pm \Omega_a\left(t\right) & E_B & \Omega_c\left(t\right)\\
\Omega_b\left(t\right) & \Omega_c\left(t\right) & E_C
\end{array}\right), \label{eq:model}
\end{eqnarray}
where the three cyclic couplings read  $\text{\ensuremath{\Omega_{a}\left(t\right)=-\mu_{a}}}\mathcal{E}_{a}\left(t\right)$, $\text{\ensuremath{\Omega_{b}\left(t\right)=-\mu_{b}}}\mathcal{E}_{b}\left(t\right)$ and $\text{\ensuremath{\Omega_{c}\left(t\right)=-\mu_{c}}}\mathcal{E}_{c}\left(t\right)$.\\ \indent
We now analyze how to achieve enantioselective excitation of a given target state from a given initial state. We assume that two enantiomers are initially in the state $|A\rangle$, and the control target can be either the excited state $|B\rangle$ or  $|C\rangle$. For the choice of $|C\rangle$ as the target, there are a direct one-photon transition $|A\rangle\leftrightarrow|C\rangle$ and an indirect two-photon transition   $|A\rangle\leftrightarrow|B\rangle\leftrightarrow|C\rangle$, which  form the closed-loop interaction scheme.  If we take  $|B\rangle$ as the target, two transition paths correspond to a direct one $|A\rangle\leftrightarrow|B\rangle$ and an indirect one $|A\rangle\leftrightarrow|C\rangle\leftrightarrow|B\rangle$. Since it remains difficult to derive an analytical solution by directly using the Hamiltonian in Eq. (\ref{eq:model}), we  use a strategy by dividing the excitation processes into two stages.
\subsection{Control conditions for ESST to $|C\rangle$}
\subsubsection{Analytical solution for a two-level system}
For the ESST to $|C\rangle$,  we assume that the coupling $\Omega_a$ is turned on at the initial time $t_{0}$ and off it at a time $t_{1}$ before the couplings $\Omega_b$ and $\Omega_c$.  Thus, the system is reduced  into a two-level system, and the corresponding Hamiltonian reads
\begin{eqnarray}
H_{1}^{L,R}\left(t\right)=\left(\begin{array}{cc}
E_A & \pm\Omega_a\left(t\right)\\
\pm\Omega_a\left(t\right) & E_B
\end{array}\right)\label{eq:model-1}.
\end{eqnarray}
Without using the rotating wave approximation, the  evolution of the system in the interaction picture can be described by using the unitary operator
\begin{equation}
U_{1}^{L,R}\left(t,t_{0}\right)=U_{1}^{L,R}\left(t_{0},t_{0}\right)-i\int_{t_{0}}^{t}dt'H_{1I}^{L,R}\left(t'\right)U_{1}^{L,R}\left(t',t_{0}\right)
\end{equation}
where $H_{1I}^{L,R}\left(t\right)=\exp\left(iH_{10}t\right)\left[\pm\Omega_{a}\left(t\right)\left(|A\rangle\langle B|+|B\rangle\langle A|\right)\right]\exp\left(-iH_{10}t\right)$
with the field-free Hamiltonian of the two-level system $H_{10}=E_{A}\left|A\right\rangle \left\langle A\right|+E_{B}\left|B\right\rangle \left\langle B\right|$. By involving the first-order the Magnus expansion \cite{pr:470:151}, the  time-dependent unitary operator $U_{1}^{L,R}\left(t,t_{0}\right)$ can be given by \cite{Shchedrin2015,shupra2019,shuprl2019,shupra2020}
\begin{eqnarray}
U_{1}^{L,R}\left(t,t_{0}\right) & = & \cos\left|\theta_{a}(t)\right|\left(\left|A\right\rangle \left\langle A\right|+\left|B\right\rangle \left\langle B\right|\right)\nonumber \\
 &  & \mp i\sin\left|\theta_{a}(t)\right|\left[\frac{\left|\theta_{a}\text{\ensuremath{\left(t\right)}}\right|}{\theta_{a}^{*}\text{\ensuremath{\left(t\right)}}}\left|B\right\rangle \left\langle A\right|+\frac{\left|\theta_{a}\text{\ensuremath{\left(t\right)}}\right|}{\theta_{a}\text{\ensuremath{\left(t\right)}}}\left|A\right\rangle \left\langle B\right|\right]
\end{eqnarray}
in terms of the complex pulse area $\theta_{a}\text{\ensuremath{\left(t\right)}}=\int_{t_{0}}^{t}\Omega_{a}\left(t'\right)\exp(i\omega_{AB}t')dt'$ with $\omega_{AB}=E_B-E_A$.
By considering the left and right-handed enantiomers initially in the ground state
$\left|A\right\rangle $, an analytic solution for the wave-function of the two-level system can be obtained by
\begin{eqnarray}
\left|\psi_{1}^{L,R}\text{\ensuremath{\left(t\right)}}\right\rangle  & = & U_{1}^{L,R}\left(t,t_{0}\right)\left|A\right\rangle \nonumber \\
 & = & \cos\left|\theta_{a}\ensuremath{\left(t\right)}\right||A\rangle \mp i\frac{\left|\theta_{a}\ensuremath{\left(t\right)}\right|}{\theta_{a}^{*}\ensuremath{\left(t\right)}}\sin\left|\theta_{a}\ensuremath{\left(t\right)}\right||B\rangle.\label{eq:wavefunction1}
\end{eqnarray}
\subsubsection{Analytical solution for a three-level system}
After the coupling $\Omega_a$ off at $t_1$, we turn on the couplings $\Omega_b$ and $\Omega_c$.  The Hamiltonian in Eq. (\ref{eq:model}) is reduced into
\begin{eqnarray}
H^{L,R}_2\left(t\right)&=&\left(\begin{array}{ccc}
E_A & 0 &  \Omega_b\left(t\right) \\
0 & E_B & \Omega_c\left(t\right)\\
\Omega_b\left(t\right) & \Omega_c\left(t\right) & E_C
\end{array}\right).
\end{eqnarray}
The corresponding time-dependent unitary operator can be given by
\begin{equation}
U^{L,R}_2(t, t_{1})=U^{L,R}_2(t_{1}, t_{1})-i\int_{t_{1}}^tdt'H_{2I}^{L,R}\left(t'\right)U^{L,R}_2(t', t_{1})
\end{equation}
where $H_{2I}^{L,R}\left(t\right)=\exp\left(iH_{20}t\right)\left[\Omega_b(t)(|A\rangle\langle C|+|C\rangle\langle C|)+\Omega_c(t)(|C\rangle\langle B|+|B\rangle\langle C|)\right]\exp\left(-iH_{20}t\right)$ with the field-free Hamiltonian of the three-level system $H_{20}=E_{A}|A\rangle\langle A|+E_{B}|B\rangle\langle B|+E_C|C\rangle\langle C|$.
By making the first-order Magnus expansion of the unitary operator $U^{L,R}_2(t, t_{1})$, the time-dependent wave function of  two enantiomers can be given by
\begin{eqnarray}
|\psi_{2}^{L,R}\ensuremath{\left(t\right)}\rangle & = & U_{2}^{L,R}\left(t,t_{1}\right)\left|\psi_{1}^{L,R}\text{\ensuremath{\left(t_{1}\right)}}\right\rangle \nonumber \\
 & = & \left[\cos\left|\theta_{a}\left(t_{1}\right)\right|\frac{\left|\theta_{c}\ensuremath{\left(t\right)}\right|^{2}+\left|\theta_{b}\ensuremath{\left(t\right)}\right|^{2}\cos\theta\ensuremath{\left(t\right)}}{\theta^{2}\ensuremath{\left(t\right)}}\mp i\sin\left|\theta_{a}\left(t_{1}\right)\zeta\ensuremath{\left(t\right)}\right|\frac{\left|\theta_{a}\ensuremath{\left(t_{1}\right)}\right|}{\theta_{a}^{*}\ensuremath{\left(t_{1}\right)}}\right]\left|A\right\rangle \nonumber \\
 &  & +\left[\cos\left|\theta_{a}\left(t_{1}\right)\right|\zeta^{*}\ensuremath{\left(t\right)}\mp i\sin\left|\theta_{a}\left(t_{1}\right)\right|\frac{\left|\theta_{a}\ensuremath{\left(t_{1}\right)}\right|}{\theta_{a}^{*}\ensuremath{\left(t_{1}\right)}}\frac{\left|\theta_{b}\ensuremath{\left(t\right)}\right|^{2}+\left|\theta_{c}\ensuremath{\left(t\right)}\right|^{2}\cos\theta\left(t\right)}{\theta^{2}\ensuremath{\left(t\right)}}\right]\left|B\right\rangle \nonumber \\
 &  & -\frac{\sin\theta\left(t\right)}{\theta\left(t\right)}\left[i\cos\left|\theta_{a}\left(t_{1}\right)\right|\theta_{b}\left(t\right)\pm\sin\left|\theta_{a}\ensuremath{\left(t_{1}\right)}\right|\theta_{c}\left(t\right)\frac{\left|\theta_{a}\ensuremath{\left(t_{1}\right)}\right|}{\theta_{a}^{*}\ensuremath{\left(t_{1}\right)}}\right]\left|C\right\rangle, \label{wfC}
\end{eqnarray}
where  $\zeta\left(t\right)=\theta_{c}\left(t\right)\theta_{b}^{*}\left(t\right)\left[\cos\theta\left(t\right)-1\right]/\theta^{2}\left(t\right)$
and $\theta\left(t\right)=\sqrt{\left|\theta_{b}\left(t\right)\right|^{2}+\left|\theta_{c}\left(t\right)\right|^{2}}$ in terms of the complex pulse areas $\theta_{b}\left(t\right)=\int_{t_{1}}^{t}\Omega_{b}\left(t'\right)\exp(i\omega_{AC}t')dt'$
and $\theta_{c}\left(t\right)=\int_{t_{1}}^{t}\Omega_{c}\left(t'\right)\exp(i\omega_{BC}t')dt'$ with the transition frequencies $\omega_{BC}=E_{C}-E_{B}$ and $\omega_{AC}=E_{C}-E_{A}$. \\ \indent

 To entirely transfer the left-handed enantiomer to the state $|C\rangle$  while keeping the right-handed one unpopulated at the final time $t_f$,  the complex pulse areas should satisfy the following two relations
\begin{eqnarray}
\left|\frac{\sin\theta\left(t_{f}\right)}{\theta\left(t_{f}\right)}\left[i\theta_{b}\left(t_{f}\right)\cos\left|\theta_{a}\left(t_{1}\right)\right|+\frac{\left|\theta_{a}\left(t_{1}\right)\right|\theta_{c}\left(t_{f}\right)}{\theta_{a}^{*}\left(t_{1}\right)}\sin\left|\theta_{a}\left(t_{1}\right)\right|\right]\right| & = & 1,\label{cd1-2}\\
\left|\frac{\sin\theta\left(t_{f}\right)}{\theta\left(t_{f}\right)}\left[i\theta_{b}\left(t_{f}\right)\cos\left|\theta_{a}\left(t_{1}\right)\right|-\frac{\left|\theta_{a}\left(t_{1}\right)\right|\theta_{c}\left(t_{f}\right)}{\theta_{a}^{*}\left(t_{1}\right)}\sin\left|\theta_{a}\ensuremath{\left(t_{1}\right)}\right|\right]\right| & = & 0,\label{cd1-1}
\end{eqnarray}
From the Eq. (\ref{cd1-1}), we can derive
\begin{equation}
\frac{\theta_{c}\left(t_{f}\right)}{\theta_{a}^{*}\left(t_{1}\right)}=\frac{i\theta_{b}\left(t_{f}\right)\cos\left|\theta_{a}\left(t_{1}\right)\right|}{\left|\theta_{a}\left(t_{1}\right)\right|\sin\left|\theta_{a}\left(t_{1}\right)\right|}. \label{cd2-1}
\end{equation}
By inserting Eq. (\ref{cd2-1}) into Eq. (\ref{cd1-2}), we can obtain a relation
\begin{equation}
4\left|\theta_{b}\left(t_{f}\right)\right|^{2}\sin^{2}\theta\left(t_{f}\right)\cos^{2}\left|\theta_{a}\left(t_{1}\right)\right|=\theta^{2}\left(t_{f}\right).\label{cd2-2}
\end{equation}
This relation can be fulfilled  when  the three control fields  satisfy the amplitude conditions
\begin{align}
\left|\theta_{b}\left(t_{f}\right)\right| & =\left|\theta_{c}\left(t_{f}\right)\right|=\frac{1}{\sqrt{2}}\left( k+\frac{1}{2}\right)\pi,(k\in N)\nonumber \\
\Big|\theta_{a}\left(t_{1}\right)\Big| & =\left(k'+\frac{1}{4}\right)\pi,(k'\in N).\label{cd3}
\end{align}
Furthermore, we insert Eq. (\ref{cd3})
into Eq. (\ref{cd2-1}) with $\theta_{b}=-\left|\theta_{b}\right|\exp\left(-i\phi_{b}\right)$, $\theta_{c}=-\left|\theta_{c}\right|\exp\left(-i\phi_{c}\right)$ and $\theta_{a}=-\left|\theta_{a}\right|\exp\left(-i\phi_{a}\right)$. We find that the three control fields satisfy the following phase condition,
\begin{equation}
\phi_{a}+\phi_{c}-\phi_{b}=\left(2l+\frac{1}{2}\right)\pi,(l\in Z).\label{cd4}
\end{equation}\\ \indent
 Similarly, we can use the same amplitude conditions as that in Eq. (\ref{cd3}) to achieve complete ESST to $|C\rangle$ of the right-handed enantiomer by using  the phase condition of $\phi_{a}+\phi_{b}-\phi_{c}=(2l-1/2)\pi,(l\in Z)$. It implies that a $\pi$ flip of the phase on one of three control fields can result in opposite ESST.  To that end, the handedness of enantiomers can be determined by measuring the population in the state $|C\rangle$.
\subsection{Control conditions for ESST to $|B \rangle$}
For ESST to $|B \rangle$, we apply the coupling $\Omega_b$ before the couplings $\Omega_a$ and $\Omega_c$, which results in a coherent superposition state of $|A\rangle$ and $|C\rangle$.  As demonstrated above by  involving the first-order Magnus expansion and further mathematical derivations, an analytic wave-function of the three-level $\Lambda$-type system can be given by
\begin{eqnarray}
|\psi_{2}^{L,R}\left(t\right)\rangle=\left[\cos\left|\theta_{b}\left(t_{1}\right)\right|\frac{\left|\theta_{c}\left(t\right)\right|^{2}+\left|\theta_{a}\left(t\right)\right|^{2}\cos\theta\left(t\right)}{\theta^{2}}\mp i\sin\left|\theta_{b}\left(t_{1}\right)\right|\xi^{*}\left(t\right)\frac{\left|\theta_{b}\left(t_{1}\right)\right|}{\theta_{b}^{*}\left(t_{1}\right)}\right] & \left|A\right\rangle \nonumber \\
-\frac{\sin\theta\left(t\right)}{\theta\left(t\right)}\left[\sin\left|\theta_{b}\left(t_{1}\right)\right|\theta_{c}^{*}\left(t\right)\frac{\left|\theta_{b}\left(t_{1}\right)\right|}{\theta_{b}^{*}\left(t_{1}\right)}\pm i\cos\left|\theta_{b}\left(t_{1}\right)\right|\theta_{a}\left(t\right)\right] & \left|B\right\rangle\nonumber \\
-\left[i\sin\left|\theta_{b}\left(t_{1}\right)\right|\frac{\left|\theta_{b}\left(t_{1}\right)\right|}{\theta_{b}^{*}\left(t_{1}\right)}\frac{\left|\theta_{a}\left(t\right)\right|^{2}+\left|\theta_{c}\left(t\right)\right|^{2}\cos\theta\left(t\right)}{\theta^{2}\left(t\right)}\mp\cos\left(\left|\theta_{b}\left(t_{1}\right)\right|\right)\xi\left(t\right)\right] & \left|C\right\rangle,
\end{eqnarray}
where $\xi\left(t\right)=\theta_{c}\left(t\right)\theta_{a}\left(t\right)\left[\cos\theta\left(t\right)-1\right]/\theta^{2}\left(t\right)$
and $\theta\left(t\right)=\sqrt{\left|\theta_{a}\left(t\right)\right|^{2}+\left|\theta_{c}\left(t\right)\right|^{2}}$.\\ \indent
To entirely transfer the left-handed enantiomer to the state $|B\rangle$ at the final time $t_f$, but  the right-handed one is not populating,  we have
\begin{eqnarray}
\left|\frac{\sin\theta\left(t_{f}\right)}{\theta\left(t_{f}\right)}\left[\sin\left|\theta_{b}\left(t_{1}\right)\right|\theta_{c}^{*}\left(t_{f}\right)\frac{\left|\theta_{b}\left(t_{1}\right)\right|}{\theta_{b}^{*}\left(t_{1}\right)}+i\cos\left|\theta_{b}\left(t_{1}\right)\right|\theta_{a}\left(t_{f}\right)\right]\right| & = & 1,\label{cdB-1}\\
\left|\frac{\sin\theta\left(t_{f}\right)}{\theta\left(t_{f}\right)}\left[\sin\left|\theta_{b}\left(t_{1}\right)\right|\theta_{c}^{*}\left(t_{f}\right)\frac{\left|\theta_{b}\left(t_{1}\right)\right|}{\theta_{b}^{*}\left(t_{1}\right)}-i\cos\left|\theta_{b}\left(t_{1}\right)\right|\theta_{a}\left(t_{f}\right)\right]\right| & = & 0.\label{cdB-2}
\end{eqnarray}
Furthermore, we can obtain the amplitude condition for the three control fields
\begin{align}
\left|\theta_{a}\left(t_{f}\right)\right| & =\left|\theta_{c}\left(t_{f}\right)\right|=\frac{1}{\sqrt{2}}\left( k+\frac{1}{2}\right)\pi, (k\in N)\nonumber \\
\Big|\theta_{b}\left(t_{1}\right)\Big| & =\left(k'+\frac{1}{4}\right)\pi,(k'\in N).\label{cd5}
\end{align}
The ESST to $|B\rangle$ of the left-handed enantiomer can be  reached by using the phase condition
\begin{equation}
\phi_{a}+\phi_{c}-\phi_{b}=\left(2l-\frac{1}{2}\right)\pi,(l\in Z). \label{cd6}
\end{equation}
The  amplitude  conditions by Eq. (\ref{cd5})  in forms are the same as Eq. (\ref{cd2-2}) with different orders. That is,  the orders of the three pulses are interchangeable, dependent on the choice of the target state.  From the phase condition in Eq.  (\ref{cd6}), we can find that a $\pi$ flip of the phase on one of three control fields  can also lead to opposite ESST.  Since our schemes that divide the closed-loop excitation schemes into two stages are different from previous works \cite{Wu2020,Torosov2020,Torosov2020-2,Wu2019,M3WM5}, which turned on the direct one-photon transition path  before the two-photon one, these amplitude and phase conditions provide an alternative way to achieve ESST  within a cyclic three-level systems in chiral molecules.   \\ \indent
Note that the amplitude conditions of $|\theta_{a/b}(t_f)|=\pi/4$ and $|\theta(t_f)|=\sqrt{|\theta_{b/a}(t_f)|^2+|\theta_c(t_f)|^2}=\pi/2$ are equivalent to  that with the use of $\pi/2$ and $\pi$ pulses,  for which a scale of $1/2$ factor comes from the definition of the complex pulse-areas without using the rotating wave approximation and the resonant excitation conditions. To show the advantage of using the complex pulse-areas, we can have  a frequency-domain analysis for the control fields
\begin{equation}
\mathcal{E}_{a/b/c}=\frac{1}{\pi}\int_{t_0}^{t_f}d\omega\mathcal{A}_{a/b/c}(\omega)e^{i\phi_{a/b/c}(\omega)}e^{i\omega t}
\end{equation}
where   $\mathcal{A}_{a/b/c}(\omega)$ and  $\phi_{a/b/c}(\omega)$ are the spectral amplitude and phase, respectively. We can find that the values of  $\theta_{a/b/c}(t_f)$ depend only on  $\mathcal{A}_{a/b/c}(\omega)$ and  $\phi_{a/b/c}(\omega)$ of three control fields at transition frequencies $\omega_{AB}$, $\omega_{AC}$ and $\omega_{BC}$.  Thus, our definitions of the complex pulse areas can also be applied to the pulsed control fields  whose  center frequencies are detuned away from the transition frequencies. In Sec. \ref{simulation}, we present simulations to examine the amplitude and phase conditions for both center-frequency resonant and detuned microwave excitation schemes.
\section{Results and discussion}\label{simulation}
We perform the numerical simulations in the cyclohexylmethanol molecules. Three rotational states of
$\left|1_{01}\right\rangle ,\left|2_{02}\right\rangle $, and $\left|2_{12}\right\rangle $ are used as
$\left|A\right\rangle, \left|B\right\rangle$, and $\left|C\right\rangle $. The transition frequencies between states are $\omega_{AB}=4720\text{MHz}, \omega_{BC}=2339\text{MHz}$,
and $\omega_{AC}=7059\text{MHz}$, and the transition dipole moments take the values of  $\mu_{a}=0.4\text{Debye}, \mu_{b}=1.2\text{Debye}$ and $\mu_{c}=0.8\text{Debye}$  \cite{Wu2020,M3WM7}. In our simulations, we take three control fields with the Gaussian profile as follows
\begin{equation}
\mathcal{E}_{a/b/c}\left(t\right)=\sqrt{\frac{2}{\pi}}\frac{A_{a/b/c}}{\tau_0}\exp{\Bigg[-\frac{(t-t_{a/b/c})^{2}}{2\tau_{0}^2}}\Bigg]\cos\left[\omega_{q/p/s}(t-t_{a/b/c})+\phi_{a/b/c}\right]. \label{field}
\end{equation}
This description of the control fields is convenient for determining the field strengths $\mathcal{E}_{a/b/c}$  for any accessible duration $\tau_0$. By choosing constant values of $A_{a/b/c}$,   we can see that the complex pulse-areas $\theta_{a/b/c}(t_f)$ with such descriptions do not depend on the duration $\tau_0$. Thus, this scheme avoids the strict limitations by the adiabatic criterion, providing a way to design fast control schemes using much shorter pulse duration than the adiabatic scenario.  For practical applications, however, we need to balance the choice of pulse duration $\tau_0$, so that unwanted transitions to neighboring energy levels could be avoided by using narrowband pulses.\\ \indent
\begin{figure*}[h]
\resizebox{0.7\textwidth}{!}{%
\includegraphics{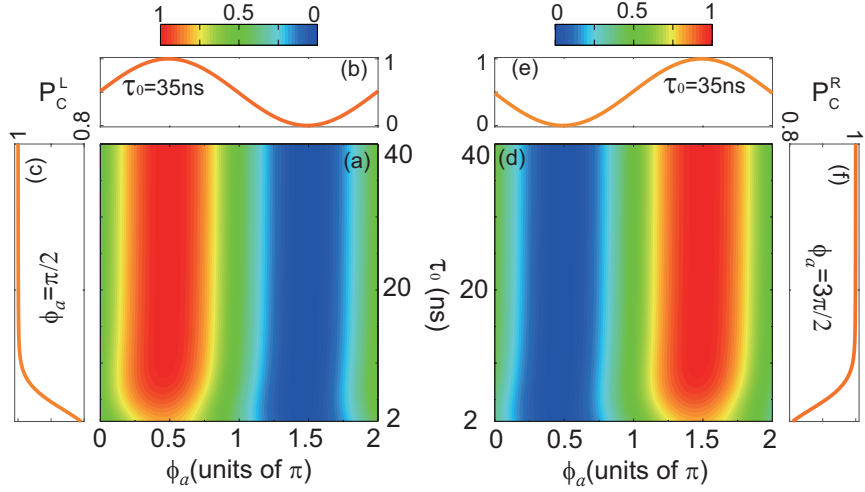} } \caption{Chiral dependence of ESST on the phase for the target state $|C\rangle$. (a) The final population of the left-handed enantiomer $P_C^L(t_f)$ versus  the duration $\tau_0$ and phase $\phi_a$ in a range of $[0, 2\pi]$, and the corresponding cut lines at (b) $\tau_0=35$ns and (c) $\phi_a=\pi/2$.  (d)-(f) The same plots as (a)-(c) for the right-handed enantiomer except for $\phi_a=3\pi/2$ in (f).}
\label{fig2}
\end{figure*}
For the cyclohexylmethanol molecules, there exists a rotational state $\left|1_{11}\right\rangle$ with the energy slightly below the state $|B\rangle$, referred  as the state $|B'\rangle$, which can be connected to the excited state $|C\rangle$ via the $a$-type transition in  $\omega_{B'C}=4484\text{MHz}$ and to the ground state $|A\rangle$ via the $c$-type transition in $\omega_{AB'}=2575\text{MHz}$. To this end, we  include this state with a four-level model to perform the numerical simulations. The corresponding field-molecule interaction Hamiltonian reads.
\begin{eqnarray}
H^{L,R}_c\left(t\right)&=&\left(\begin{array}{cccc}
0 & \Omega^{'}_{c}(t) & \pm\Omega_a(t) &\Omega_b(t)\\
\Omega^{'}_{c}(t) & 0 & 0 &\pm\Omega_a^{'}(t)\\
\pm\Omega_a(t) & 0 & 0 & \Omega_c(t)\\
\Omega_b(t) & \pm\Omega_a^{'}(t) & \Omega_c(t) & 0
\end{array}\right), \label{eq:model4}
\end{eqnarray}
where we take the additional couplings $\Omega_a^{'}(t)=-\mu'_{a}\mathcal{E}_a(t)$ and $\Omega_c^{'}(t)=-\mu'_{c}\mathcal{E}_c(t)$ with $\mu'_a=\mu_a$ and $\mu'_c=\mu_c$ in our simulations. The time-dependent unitary operator in the interaction picture can be numerically computed by
\begin{equation}
U^{L,R}(t, t_{0})=U^{L,R}(t_{0}, t_{0})-i\int_{t_{0}}^tdt'H_{I}^{L,R}\left(t'\right)U^{L,R}(t', t_{0}),
\end{equation}
where $U^{L,R}(t_{0},t_0)=\mathbb{I}$ and $H_{I}^{L,R}\left(t\right)=\exp\left(iH_{0}t\right)[H^{L,R}_c\left(t\right)]\exp\left(-iH_{0}t\right)$ with the field-free Hamiltonian $H_0=\sum_{X=A}^CE_X|X\rangle\langle X|$. By projecting the unitary operator $U^{L,R}(t, t_{0})$ onto the initial state $|A\rangle$, we can obtain the time-dependent wave function of the system $|\psi^{L,R}(t)\rangle=U^{L,R}(t, t_{0})|A\rangle$ without using the first-order Magnus expansion. Thus, the time-dependent population in the  state $|X\rangle$  can be calculated by $P^{L/R}_{X}(t)=\left|\langle X|\psi^{L, R}(t)\rangle\right|^2$ with $X=A, B', B, C$.
%We define a parameter $S=P_L(t_f)-P_R(t_f)$ to evaluate the enantioselective excitation of the target state $|F\rangle$ in enantiomers, i.e.,  1 for the left handedness or -1 for the right handedness.
\subsection{ESST to $|C\rangle$}
\begin{figure}[h]
\resizebox{0.7\textwidth}{!}{%
\includegraphics{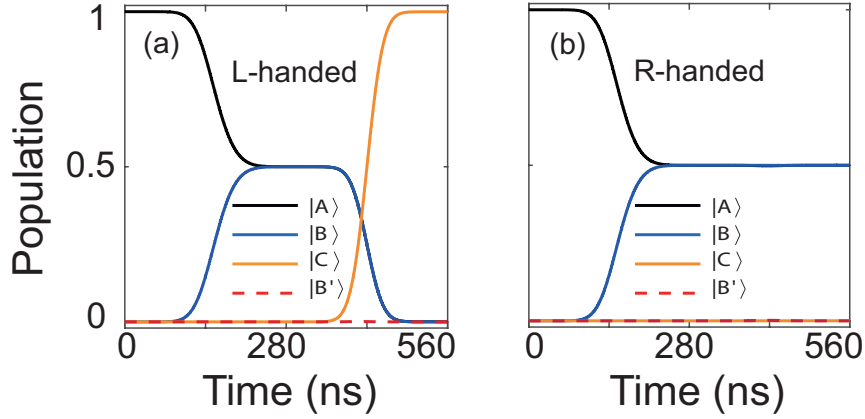} }\caption{The time-dependent populations of four rotational states for ESST to the state  $|C\rangle$.   The simulations for (a)  the left-handed and (b) right-handed enantiomers with the duration  $\tau_0=35$ns and phase $\phi_a=\pi/2$. }
\label{fig3}
\end{figure}
 For the target state $|C\rangle$, we set the parameters $A_a=\pi/(4\mu_a)$, $A_b=\pi/(2\sqrt{2}\mu_b)$, and $A_c=\pi/(2\sqrt{2}\mu_c)$.  It is easily to verify that the three control fields defined by Eq. (\ref{field}) with different values of $\tau_0$ exactly satisfy the amplitude conditions in Eq. (\ref{cd3}) at transition frequencies by fixing the center frequencies $\omega_a=\omega_{AB}$, $\omega_b=\omega_{AC}$ and $\omega_c=\omega_{BC}$. Figure \ref{fig2} shows the results of $P_C^{L, R}(t_f)$ versus $\tau_0$ and $\phi_a$ with $\phi_b=\phi_c=0$. As expected,  the ESST to the state $|C\rangle$ appears and  depends highly on the phase values of $\phi_a$. The fidelity of $P_C^{L, R}(t_f)>0.999$  can be reached for $\tau_0>35$ ns, indicating that the unwanted transition to  the neighboring state $|B'\rangle$ can be ignored. Figures \ref{fig2} (b) and (e) plot the dependence of $P_C^{L, R}(t_f)$ on the phase $\phi_a$ for the case of $\tau_0=35$ ns. There is no ESST  at $\phi_a=0$ and $\pi$. The entire ESST  to the left-handed molecule occurs at $\phi_a=\pi/2$ and a phase change to $\phi_a=3\pi/2$ results in an opposite transfer to  the right-handed one. The similar features can be observed by changing the values of $\phi_b$ ( or $\phi_c$) while choosing $\phi_a=\phi_c=0$ ( or $\phi_a=\phi_b=0$). These results are in good agreement with the theoretical predication by the phase conditions as well as previous M3WM experiments. \\ \indent
 \begin{figure}[h]
\resizebox{0.7\textwidth}{!}{%
\includegraphics{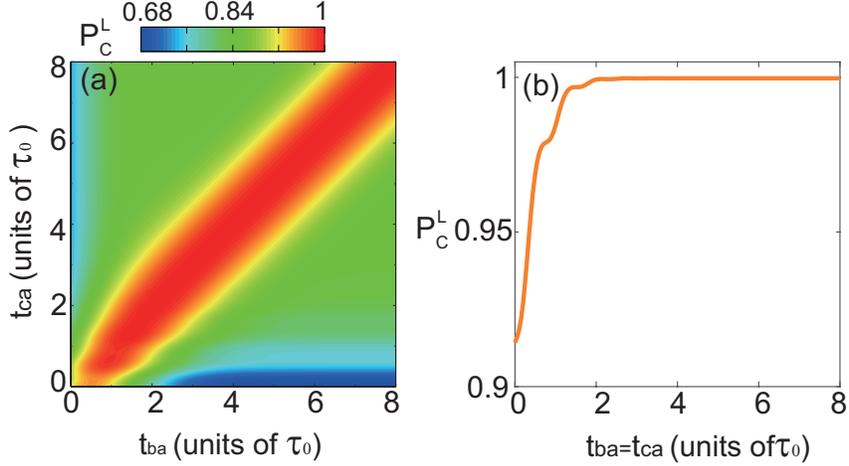} } \caption{The dependence of ESST  on the time delays of control fields for the target state $|C\rangle$. (a) The final population of the left-handed enantiomer $P_C^{L}(t_f)$ versus the time delays $t_{ba}=t_b-t_a$ and $t_{ca}=t_c-t_a$. (b) The cut line plot of $P_C^{L}(t_f)$ along $t_{ba}=t_{ca}$. }
\label{fig4}
\end{figure}
To visualize the underlying quantum state transfer mechanism, Fig.\ref{fig3} shows the time-dependent populations of the four states induced by the control fields for the cases of  $\tau_0=35$ ns and $\phi_a=\pi/2$. For the two enantiomers,   there are no visible populations in the state $|B'\rangle$ during the whole process. The four-level system is equivalent to the present closed-loop three-level model with the used pulse parameters. The control field $\mathcal{E}_a(t)$ with the pulse-areas $\theta_a(t_1)=\pi/4$ drives the  system to  the maximal coherent superposition of $|A\rangle$ and $|B\rangle$ with $P_A^{L,R}(t_1)=P_B^{L,R}(t_1)=0.5$ for both  enantiomers. Due to the sign difference of the transition from $|A\rangle$ to $|B\rangle$, $\mathcal{E}_a(t)$ with a phase $\phi_a=\pi/2$ will result in the phase of the state $|B\rangle$ in $0$ and $\pi$ for the left- and right-handedness, respectively, as described by Eq. (\ref{wfC}).  For the left-handedness, the transition path from $|A\rangle$ to $|C\rangle$  induced by  $\mathcal{E}_b(t)$ will constructively with the path from $|B\rangle$ to $|C\rangle$ by  $\mathcal{E}_c(t)$, leading to entire ESST to $|C\rangle$. For the right-handedness, however, the two paths  are destructive, which keeps the molecules in the states $|A\rangle$ and $|C\rangle$, as shown in Fig. \ref{fig3} (a) and (b).\\ \indent
As can be seen from our theoretical derivations, we divide the closed-loop excitation scheme into two time-separated stages. To see whether the amplitude and phase conditions can be applied to the overlapped cases, as an example, Fig. \ref{fig4} plots the landscape of $P_C^L(t_f)$ with respect to the time delays $t_{ba}=t_b-t_a$ and $t_{ca}=t_c-t_a$ while fixing the center time $t_a$ unchanged. $P_C^L(t_f)$ strongly depends on the overlap between two control fields of the second stage. Interestingly,   $P_C^L(t_f)$ remains the value of $P_C^L(t_f)>0.999$ for $t_{ba}=t_{ca}>2\tau_0$ when  $\mathcal{E}_b(t)$ and $\mathcal{E}_c(t)$ are turned on  simultaneously with  $t_{ba}=t_{ca}$. Even the three control fields are applied without any delays, high fidelity  of $P_C^L(t_f)>0.90$ holds, as shown in Fig. \ref{fig4} (b). The  phenomena can also be observed for the right-handedness (not shown here).
\subsection{ESST to $|B\rangle$}
\begin{figure}[h]
\resizebox{0.7\textwidth}{!}{%
\includegraphics{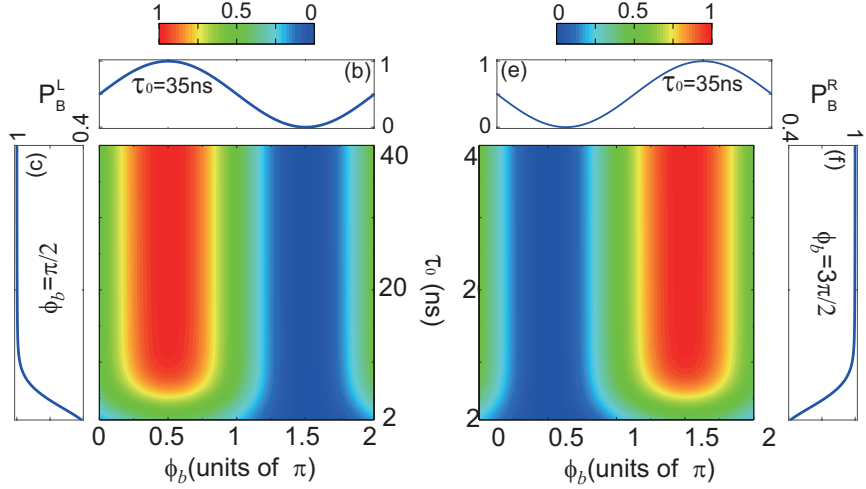} }\caption{Chiral dependence of ESST on the phase for the target state $|B\rangle$. (a) The final population of the left-handed enantiomer $P_C^L(t_f)$ versus  the duration $\tau_0$ and phase $\phi_b$ in a range of $[0, 2\pi]$, and the corresponding cut lines at (b) $\tau_0=35$ ns and (c) $\phi_b=\pi/2$.  (d)-(f) The same plots as (a)-(c) for the right-handed enantiomer except for $\phi_a=3\pi/2$ in (f).}
\label{fig5}
\end{figure}
Figure \ref{fig5} examines the same simulations as  Fig. \ref{fig2} but for  the target $|B\rangle$ with $\phi_a=\phi_c=0$. In our simulations, we choose the parameters $A_b=\pi/(4\mu_a)$, $A_a=\pi/(2\sqrt{2}\mu_b)$, and $A_c=\pi/(2\sqrt{2}\mu_c)$ so as to the all fields satisfy the amplitude conditions. The influence of the state $|B'\rangle$  looks more visible than that in Fig. \ref{fig2} in the short duration regime, which becomes rather weak by increasing duration $\tau_0$.  The final population $P_B^{L,R}(t_f)$ can also reach high fidelity for $\tau_0>35$ns. As demonstrated in Fig. \ref{fig2}, the landscape of $P_B^{L,R}(t_f)$ exhibits a chiral symmetry with respect to the phase $\phi_b$, for which the control field $\mathcal{E}_b(t)$ with $\phi_b=\pi/2$ leads to entire ESST to $|B\rangle$ of the left-handedness. For the right-handedness,  however, it requires to  $\phi_b=3\pi/2$. This dependence of $P_B^{L,R}(t_f)$ on the phase is  consistent with the theoretical predication.\\ \indent
 \begin{figure}[h]
\resizebox{0.7\textwidth}{!}{%
\includegraphics{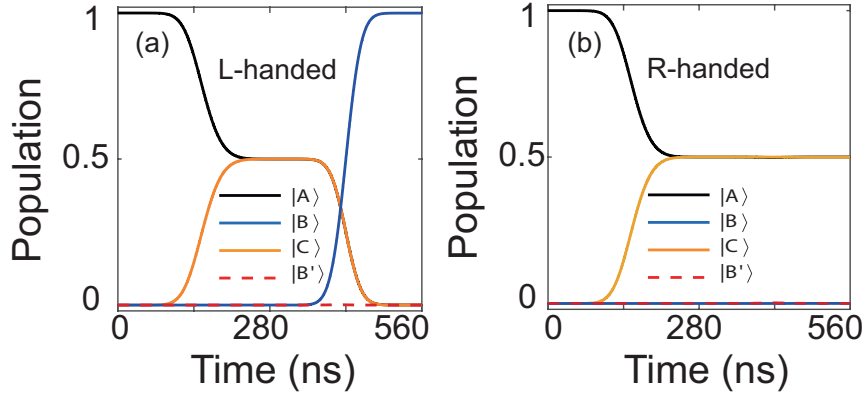} }\caption{The time-dependent populations of four rotational states for ESST to the state  $|B\rangle$.   The simulations for (a)  the left-handed and (b) right-handed enantiomers with the duration  $\tau_0=35$ns and phase $\phi_b=\pi/2$. }
\label{fig6}
\end{figure}
Figure \ref{fig6} plots the time-dependent populations of the system with  $\tau_0=35$ and $\phi_b=\pi/2$. Since the transition moments $\mu_b$ are identical for the two enantiomers without a difference of sign,   $\mathcal{E}_b(t)$ plays the same role in the first stage, generating the same maximal coherent superposition of $|A\rangle$ and $|C\rangle$. The constructive or destructive interference that depends on $\mu_a$ occurs between the transition paths from $|A\rangle$ and $|C\rangle$ to $|B\rangle$. As a result, the left-handed enantiomer is fully transferred to the state $|B\rangle$, whereas the right-handed one is still in the coherent states $|A\rangle$ and $|B\rangle$ at end of three pulses. \\ \indent
 \begin{figure}[h]
\resizebox{0.7\textwidth}{!}{%
\includegraphics{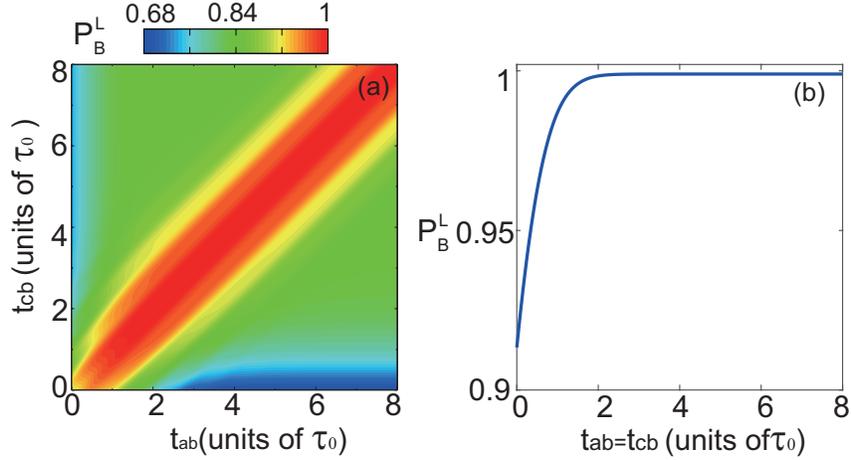} } \caption{The dependence of ESST  on the time delays of control fields for the target state $|B\rangle$. (a) The final population of the left-handed enantiomer $P_B^{L}(t_f)$ versus the time delays $t_{ba}=t_b-t_a$ and $t_{ca}=t_c-t_a$. (b) The cut line plot of $P_B^{L}(t_f)$ along $t_{ba}=t_{ca}$. }
\label{fig7}
\end{figure}
To see the robustness of the scheme on the time delays, Fig. \ref{fig7} examines the dependence of $P_B^L(t_f)$ on the time delays $t_{ab}=t_a-t_b$ and $t_{cb}=t_c-t_b$. Similar behaviors can be observed, indicating that both excitation schemes do not require strict separations between the control fields. The identical delays of the second stage fields are beneficial for the control. As a result, the amplitude and phase conditions can also be used for the overlapped control fields, leading to the high selectivity of handedness.
\subsection{ESST with center frequency-detuned pulses}
Finally, we examine the amplitude and phase conditions of control fields whose center frequencies $\omega_{a/b/c}$ are not exactly resonant with the transition center frequencies $\omega_{AB}$, $\omega_{BC}$ and $\omega_{AC}$. As can be seen from the definitions of the complex pulse areas $\theta_{a/b/c}(t_f)$, when the center frequencies are detuned away from resonances, the values of $|\theta_{a/b/c}(t_f)|$ will be decreased while keeping the parameters $A_{a/b/c}$ unchanged as used in resonant cases. We can increase the values of $A_{a/b/c}$ to revive the values of $\theta_{a/b/c}(t_f)$ at the transition frequencies so as to satisfy the amplitude conditions. That is, ESST in principle could be achieved by using the center-frequency-detuned pulses, as long as they satisfy the amplitude and phase conditions  $\mathcal{E}_{a/b/c}$ and $\phi_{a/b/c}$ at transition frequencies.\\ \indent
\begin{figure}[h]
\resizebox{0.7\textwidth}{!}{%
\includegraphics{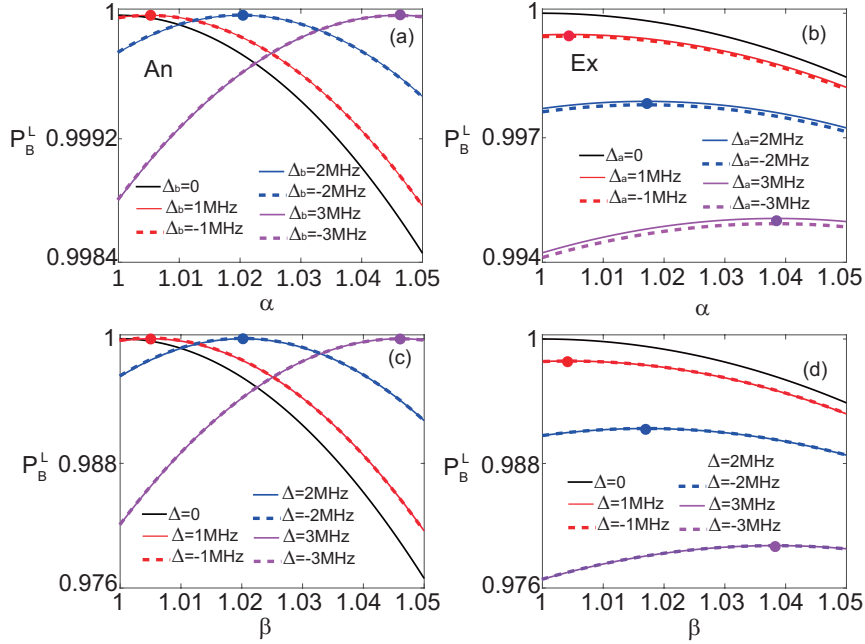} } \caption{The dependence of ESST on  the detunings for the target state $|B\rangle$. and (b) The final population $P_{B}^{L}(t_f)$ versus the scale factor $\alpha$ of the coupling $\Omega_b(t)$ for different values of detuning $\Delta_b=\omega_b-\omega_{AC} $. (c) and (d) The final population $P_{B}^{L}(t_f)$ versus the scale factor $\beta$ of the couplings $\Omega_a(t)$ and $\Omega_c(t)$ by taking $\Delta_a=\omega_a-\omega_{AB}=\Delta_c=\omega_c-\omega_{BC}=\Delta$. The analytical simulations (left panels) are compared with the exact results (right panels). }
\label{fig8}
\end{figure}
Figure \ref{fig8} shows the dependence of  $P_{B}^{L}(t_f)$ on the detunings, for which the analytical simulations (left panels) are compared with the exact ones (right panels). We perform the simulations in Figs. \ref{fig8} (a) and (b) to calculate  $P_{B}^{L}(t_f)$ for different values of  $\Delta_b=\omega_b-\omega_{AC}$ while scaling the coupling $\Omega_b(t)$ with a factor $\alpha$, for which the center frequencies $\omega_b$ and $\omega_c$ are fixed at the resonant conditions. The simulations in Figs. \ref{fig8} (c) and (d) are accomplished with different detunings $\Delta_a=\omega_a-\omega_{AB}$ and $\Delta_c=\omega_c-\omega_{BC}$ by scaling the couplings $\Omega_a(t)$ and $\Omega_c(t)$ with a factor $\beta$, for which we set  $\Delta_a=\Delta_
c=\Delta$ while fixing $\Delta_b=0$.  For both analytical simulations, we can see that the detunings decrease $P_{B}^{L}(t_f)$. By increasing the strengths of the control fields, the maximal value of $P_{B}^{L}(t_f)$  can be revived to the same level as the resonant excitation, as shown in Figs. \ref{fig8} (a) and (c).  For the exact simulations, however,  the maximal values can be increased but below the theoretical maximum.  The differences can be attributed to the influence of high-order Magnus expansion terms, which are ignored in the analytical model. We also observe similar results for target state $|B\rangle$ (not shown here). Thus, the center-frequency detuned excitations with small detunings are also allowed by applying the corresponding amplitude and phase conditions at transition frequencies, whereas the larger detunings will  reduce the ESST efficiency due to the optical processes via high-order high-order Magnus terms.
\section{Conclusion}\label{con}
We presented a general pulse-areas theorem analysis to explore ESST within a closed-loop three-level system. We considered three rotational states cyclically connected by the $a$-type, $b$-type, and $c$-type components of the transition dipole molecules. Using a strategy that separates the closed-loop excitation into two stages, we derived the amplitude and phase conditions for designing three linearly polarized microwave pulses to generate ESST to different targets. The two-stage strategy we used differs from previous schemes that turned on the direct one-photon transition from the initial state to the target state before the indirect two-photon one. Our schemes firstly switched on one control field involved in the two-photon path by generating maximal coherent supposition between the initial and intermediate states. We examined this three-level pulse-areas theorem in the cyclohexylmethanol molecules and analyzed its applications with both center-frequency resonant and detuned pulse sequences. For the latter, small detunings on the center frequencies of the control pulses would be expected to reduce the influence of high-order Magnus expansion terms. It opens a fundamental question of whether one can design fast and robust quantum control schemes against center-frequency detunings. To that end, optimal and robust control methods combined with artificial intelligence algorithms  could be used to search for shaped control pulses subject to multiple constraints \cite{shupra2019,QOCT2,shu6,IEEEshu,Yang2020}.
\begin{acknowledgements}
This work was supported by the National Natural Science
Foundations of China (NSFC) under Grant No. 61973317.  Y. G. is partially supported by
the Scientific Research Fund of Hunan Provincial Education Department under Grant No. 20A025, Changsha Municipal Natural Science Foundation under Grant No. kq2007001,  the Opening Project of Key Laboratory of Low Dimensional Quantum Structures and Quantum Control of the Ministry of Education under Grant No. QSQC1905, and the Open Research Fund of Hunan Provincial Key Laboratory of Flexible Electronic Materials
Genome Engineering under Grant No. 202009.
\end{acknowledgements}

\end{document}